\documentclass[]{spie} 
 
\usepackage{amsmath,amsfonts,amssymb}
\usepackage{graphicx}
\usepackage[colorlinks=true, allcolors=blue]{hyperref}
\usepackage{multirow}
\usepackage[symbol]{footmisc}

\title{\Large\bfseries Dual Purpose Lyot Coronagraph Masks for Simultaneous High-Contrast Imaging and High-Resolution Wavefront Sensing}

\author[a]{G. Ruane}
\author[a]{J. K. Wallace}
\author[a]{A. J. E. Riggs}
\author[a]{T. Wenger}
\author[a]{M. Bagheri}
\author[a]{J. Jewell}
\author[a]{N. Raouf}
\author[a]{G. Allan}
\author[a]{C. Mejia~Prada}
\author[a]{M. Noyes}
\author[a]{A. B. Walter}
\affil[a]{Jet Propulsion Laboratory, California Institute of Technology, 4800 Oak Grove Dr., Pasadena, CA 91109}

\authorinfo{Send correspondence to gruane@jpl.nasa.gov.}

\begin{document} 
\maketitle

\begin{abstract}
Directly imaging Earth-sized exoplanets with a visible-light coronagraph instrument on a space telescope will require a system that can achieve $\sim10^{-10}$ raw contrast and maintain it for the duration of observations (on the order of hours or more). We are designing, manufacturing, and testing Dual Purpose Lyot coronagraph (DPLC) masks that allow for simultaneous wavefront sensing and control using out-of-band light to maintain high contrast in the science focal plane. Our initial design uses a tiered metallic focal plane occulter to suppress starlight in the transmitted coronagraph channel and a dichroic-coated substrate to reflect out-of-band light to a wavefront sensing camera. The occulter design introduces a phase shift such that the reflected channel is a Zernike wavefront sensor. The dichroic coating allows higher-order wavefront errors to be detected which is especially critical for compensating for residual drifts from an actively-controlled segmented primary mirror. A second-generation design concept includes a metasurface to create polarization-dependent phase shifts in the reflected beam, which has several advantages including an extended dynamic range. We will present the focal plane mask designs, characterization, and initial testing at NASA’s High Contrast Imaging Testbed (HCIT) facility.
\end{abstract}

\keywords{high contrast imaging, coronagraphs, exoplanets}

\section{Introduction}
\label{sec:intro} 

As part of NASA's response to the Astro2020 Decadal Survey\cite{Astro2020}, a technology maturation program is underway to prepare for the future Habitable Worlds Observatory (HWO), which is largely inspired by the LUVOIR\cite{LUVOIR_finalReport} and HabEx\cite{HabEx_finalReport} mission concepts. HWO is a 6-meter space telescope concept that aims to directly image and spectrally characterize Earth-sized exoplanets in the habitable zone of Solar-type (FGK) stars. A coronagraph instrument designed for this purpose has two primary tasks: (1)~to achieve raw contrasts on the order of $10^{-10}$ in specific fields-of-view and spectral bandwidths and (2)~to maintain that raw contrast for the timescales of typical observations (hours to weeks).

In terms of raw contrast, the best performing coronagraph designs to date are Lyot coronagraphs, which have been demonstrated to achieve contrasts of $\sim4\times10^{-10}$ in a 10\% bandwidth with central wavelength of $\lambda_0$~=~550~nm averaged over a 3-9 $\lambda_0/D$ annular region around the star\cite{Seo2019}, where $D$ is the pupil diameter. A variant of Lyot coronagraphs, known as Hybrid Lyot Coronagraphs (HLCs)\cite{Trauger2012}, are also used as part of the Roman Space Telescope Coronagraph Instrument\cite{Riggs2021_SPIE_CGI}. HLCs have a focal plane mask that consists of a metallic occulter approximately 6~$\lambda/D$ in diameter on a glass substrate with a dielectric pattern on the front of the metal region. The metal spot in this design blocks starlight from passing to the science camera and reflects it towards the low-order wavefront sensor (LOWFS) camera\cite{Shi2016_JATIS}. Since the reflected beam is spatially filtered by the occulter, the LOWFS senses spatial frequencies up to approximately 3 cycles across the pupil. 

The current HWO working design is a segmented telescope, which has important implications on the design of the coronagraph masks and the wavefront sensor (WFS). In the following, we present a ``Dual Purpose" Lyot coronagraph (DPLC) concept that is specifically designed to allow sensing of higher order aberrations. The DPLC is a Lyot coronagraph that reflects starlight outside of the coronagraph science filter to a high-order (beyond 3 cycles per pupil diameter) WFS camera. The underlying strategy is to minimally modify the best performing coronagraph to date to also serve the role of an out-of-band wavefront sensor operating simultaneously with coronagraph observations. The DPLC allows for wavefront corrections during coronagraph observations, provides sufficient resolution to correct for optical path differences due to residual segment motions, minimizes non-common path aberrations, and maximizes the sensitivity to small changes in the wavefront incident on the coronagraph focal plane mask. In this paper, we discuss the system-level design considerations for the DPLC (Section~\ref{sec:systemlevel}) and present our first generation mask design (Section~\ref{sec:maskdesign}), performance modeling (Section~\ref{sec:performance}), our initial laboratory results (Section~\ref{sec:results}), and conclusions (Section~\ref{sec:conclusions}). 

\section{System-level design considerations} 
\label{sec:systemlevel} 

\begin{figure}
    \centering
    \includegraphics[width=\linewidth]{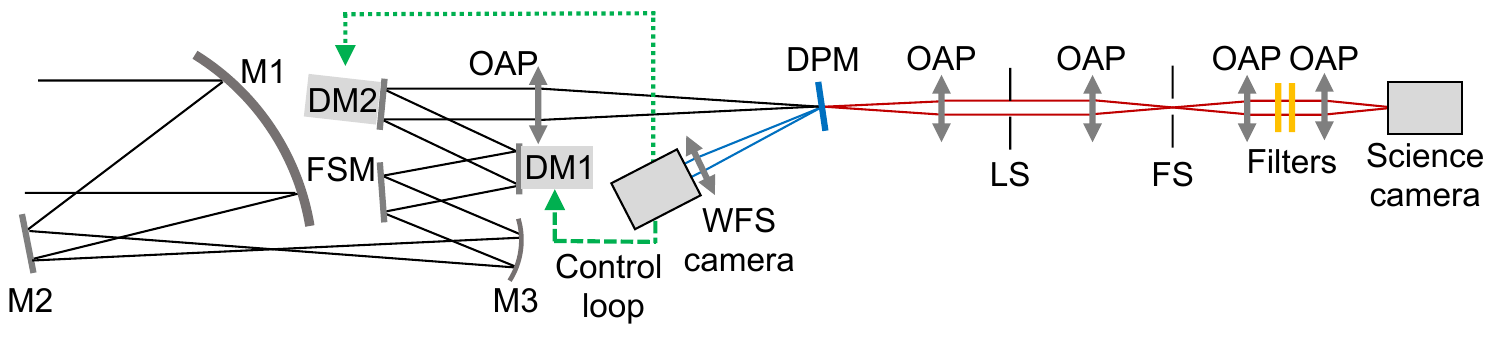}
    \caption{System-level schematic (not to scale) of a possible off-axis space telescope and coronagraph instrument. The telescope (shown as M1, M2, and M3) delivers the light from a stellar system to the coronagraph instrument. The coronagraph is made up of a fast steering mirror (FSM), deformable mirrors (DM1 and DM2), off-axis parabolic (OAP) mirrors, a dual purpose mask (DPM) in a focal plane, a Lyot stop (LS), a field stop (FS), optical filters, and a science camera. The light reflected from the DPM goes to one or more wavefront sensor (WFS) cameras. Feedback from the WFS is used to control the wavefront using one of more deformable mirrors in a closed-loop fashion. }
    \label{fig:schematic}
\end{figure}

Figure \ref{fig:schematic} shows a system-level schematic of a space telescope with a coronagraph instrument designed for exoplanet imaging. We assume there is one or more deformable mirror (DM), a dual purpose mask (DPM) in the focal plane, one or more WFS cameras that receives light reflected from the DPM, and a science camera for imaging exoplanets in the high-contrast image. The coronagraph produces a region of high contrast (a.k.a. ``dark hole" or ``dark zone") on the science camera over a specified field-of-view and wavelength range via iterative focal-plane wavefront sensing and control\cite{Groff2015}. Then this state is maintained using a closed control loop that uses feedback from the WFS camera(s) to correct the DM settings to compensate for changes in the upstream electric field.  

The coronagraph instrument on HWO will be required to achieve $\sim$10$^{-10}$ raw contrasts in $\geq$10\% spectral bandwidths covering at least the visible range.  To date, broadband laboratory demonstrations that achieve contrasts better than 10$^{-9}$ have been limited to simple Lyot coronagraphs\cite{Seo2019}. Recent results with vector vortex coronagraphs\cite{Ruane2022} achieved approximately 10$^{-9}$. The drawbacks of Lyot coronagraphs compared to vortex coronagraphs include relatively high sensitivity to low-order aberrations, lower throughput, and limited instantaneous fields-of-view (i.e. dark hole sizes). On the other hand, the contrasts achieved on coronagraph testbeds with simple Lyot coronagraphs are consistently within a factor of a few of HWO requirements. Thus, Lyot coronagraphs remain a primary option until further maturation of other technologies with the potential for improved performance. Given the fundamental need for high contrast ($\sim$10$^{-10}$), we adopt a basic Lyot coronagraph design in this work. Future work will consider how other coronagraph technologies that achieve the required contrast levels can be modified to allow high-order sensing in a similar fashion to the DPMs presented here. 

Like the Coronagraph Instrument on the Roman Space Telescope (RST), we adopt a WFS design that uses the starlight reflected from the focal plane mask in the coronagraph. There are a few important advantages to this approach. The first is that it removes the need for an upstream beamsplitter in the stellar beam path, which is likely to introduce ghost reflections and other aberrations (chromatic, polarization, etc) that negatively impact the raw contrast. Given the challenges encountered in past laboratory demonstrations of coronagraphs\cite{Seo2017,Seo2019,Ruane2022}, our strategy is to preemptively avoid potential contributors to additional unwanted stellar leakage. 

The second advantage of a reflective focal plane mask is that it minimizes the non-common path aberrations with respect to the starlight in the science band. In Lyot coronagraphs, the raw contrast in the image plane is most sensitive to wavefront changes upstream of the occulting mask. Wavefront changes that occur downstream of the coronagraph masks have orders of magnitude lower impact. For example, methods that measure the end-to-end wavefront to the science camera and use the DM to remove errors based on that measurement can negatively impact raw contrast by injecting the wavefront errors originating downstream of the coronagraph masks to the plane of the DM. Using a beamsplitter in the system upstream of the coronagraph masks may also be favorable in this regard, but the ideal configuration in terms of achieving the contrast goals is to have the focal plane mask substrate as the first transmissive optic in the stellar beam. 

The third advantage of using the focal plane mask reflection is that the WFS operates simultaneously with high-contrast imaging observations. Indeed, RST has a fast steering mirror and DM that apply corrections during observations using low-order WFS measurements. This improves raw contrast and potentially extends the possible duration of observations. The DPLC aims to enable similar functionality.

Sensing residual phasing errors on a segmented telescope will require a WFS that is sensitive to higher spatial frequencies than that of the RST coronagraph's WFS. HWO could have 5 or more mirror segments across the primary. Segment-to-segment errors in the form of rigid body motions will require at least a few wavefront samples per segmented mirror area, and at least two across to measure relative tilts. As a result, the focal plane reflecting region needs to be more than $\sim$20~$\lambda/D$ in diameter to detect and correct such errors. If the WFS needs to measure DM instabilities at the actuator level, many more samples (potentially 48 to 128 actuators across) and a much larger reflecting region (respectively, $>$24 to 64~$\lambda/D$) will be required. However, a Lyot coronagraph occulter is typically 6~$\lambda/D$ in diameter. This implies that the region outside of the occulter also needs to be reflective at the wavefront sensing wavelengths. 

To stabilize the raw contrast to levels $<$10$^{-10}$, the wavefront changes need to be $\lesssim$10~pm~RMS at all timescales ideally up to the duration of an observation\cite{JuanolaParramon2022}. Thus, the WFS should be designed to sense picometer-level OPD changes at the required spatial resolutions in the shortest amount of time possible. A Zernike (a.k.a. phase-contrast) WFS\cite{Dicke1975,Vorontsov2000,Bloemhof2004} has been demonstrated to provide near ideal sensitivity to such wavefront changes in a coronagraph instrument and thus is our preferred sensing method\cite{Guyon2005,Steeves2020,Ruane2020_JATIS}. 

\begin{figure}
    \centering
    \includegraphics[width=0.8\linewidth]{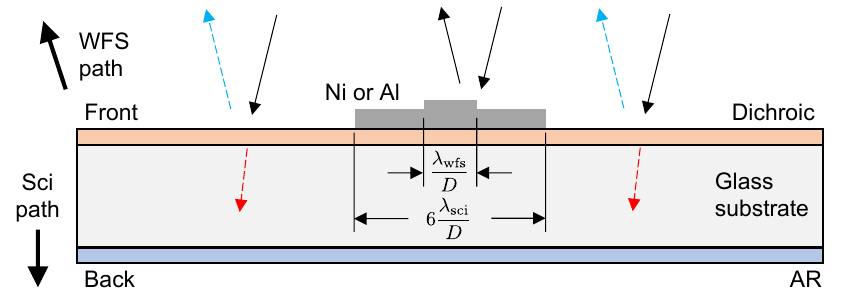}
    \caption{Diagram (not to scale) of a dual-purpose Lyot coronagraph (DPLC) focal plane mask.  }
    \label{fig:fpm_design}
\end{figure}

\section{Focal plane mask design} 
\label{sec:maskdesign}

Given the design considerations laid out in the previous section, we arrived at the notional focal plane mask concept shown in Fig.~\ref{fig:fpm_design}. The mask consists of a glass substrate that is typically 5-6~mm thick and 25.4~mm in diameter. A two-tiered metallic occulter made from $\sim$100~nm thick Nickel (Ni) or Aluminium (Al) reflects light from the star (required transmission $\lesssim$10$^{-4}$) in an approximately 6~$\lambda_\text{sci}/D$ diameter region, where $\lambda_\text{sci}$ is the wavelength in the science band. The residual transmitted beam continues through the coronagraph to the science camera (Sci path) and the reflection from the whole device becomes the WFS path. While the reflection from the metallic region alone provides low-order spatial information at the WFS, introducing a dichroic coating to the substrate reflects the out-of-band light containing high-spatial frequency information. Both the dichroic and anti-reflective (AR) coating on the backside have high transmission over the science wavelength range to prevent secondary reflections that degrade contrast at the science camera. The result is a coronagraph mask that works similarly to that of the best performing coronagraphs to date\cite{Seo2019}, while simultaneously providing light for high-order wavefront sensing with our previously demonstrated technique\cite{Ruane2020_JATIS} but instead using the out-of-band wavelengths reflected from the dichroic. 

\begin{table}[t]
\caption{Parameters for the first generation DPLC design.} 
\label{tab:fpm_design}
\begin{center}       
\begin{tabular}{|c|c|} 
\hline
\rule[-1ex]{0pt}{3.5ex}  WFS wavelength range & 500-550~nm \\
\hline
\rule[-1ex]{0pt}{3.5ex}  Sci wavelength range & 626-700~nm \\
\hline
\rule[-1ex]{0pt}{3.5ex}  Occulter material & Aluminum \\ 
\hline
\rule[-1ex]{0pt}{3.5ex}  Full occulter diameter & 124.8~$\mu$m \\ 
\hline
\rule[-1ex]{0pt}{3.5ex}  Phase shifter diameter & 16.8~$\mu$m \\ 
\hline
\rule[-1ex]{0pt}{3.5ex}  Scalar phase shifter height & 65.6~nm \\ 
\hline
\rule[-1ex]{0pt}{3.5ex}  Occulter base thickness & $>$70~nm \\ 
\hline
\rule[-1ex]{0pt}{3.5ex}  Angle of incidence & 5.5$^\circ$ \\ 
\hline
\end{tabular}
\end{center}
\end{table}

The top tier of the occulter in Fig.~\ref{fig:fpm_design} is specially designed to enable wavefront sensing. The light reflected from the central region $\sim\lambda_\text{wfs}/D$ in diameter, where $\lambda_\text{wfs}$ is the WFS wavelength, is phase shifted by approximately $\lambda_\text{wfs}/4$. The light in the WFS band that reflects from the dichroic combines with the light reflected from this scalar phase shifter and the other parts of the occulter to allow high-order wavefront sensing. Meanwhile, the light in the science band that reflects from the metallic region can also be used for low-order wavefront sensing. The detailed parameters for our first generation mask design are given in Table~\ref{tab:fpm_design}.

\begin{table}[t]
\caption{Parameters for the dichroic designs using alternating layers of HfO$_2$ and SiO$_2$. $^\dagger$Total depth of the coating.} 
\label{tab:coating_parameters}
\begin{center}       
\begin{tabular}{|c|c|c|c|c|c|c|c|c|c|c|} 
\hline
\rule[-1ex]{0pt}{3.5ex}  Coating & \# layers & Depth$^\dagger$ ($\mu$m) & \multicolumn{4}{|p{4cm}|}{Mean reflectance (\%)}  & \multicolumn{2}{|p{2cm}|}{Mean phase} & \multicolumn{2}{|p{2cm}|}{Phase range}  \\
\hline
\rule[-1ex]{0pt}{3.5ex}   &  &  & \multicolumn{2}{|c|}{WFS band} & \multicolumn{2}{|c|}{Sci band}  & \multicolumn{2}{|c|}{WFS band} & \multicolumn{2}{|c|}{WFS band}  \\
\hline
\rule[-1ex]{0pt}{3.5ex} & & & $\perp$ & $\parallel$ & $\perp$ & $\parallel$ & $\perp$ & $\parallel$ & $\perp$ & $\parallel$  \\
\hline
\rule[-1ex]{0pt}{3.5ex} A & 17 & 1.39 & 97.3 & 97.4 & 1.00 & 1.04 & -0.80 & -0.81 & 0.24 & 0.24 \\
\hline
\rule[-1ex]{0pt}{3.5ex} B & 23 & 1.80 & 98.4 & 98.4 & 0.64 & 0.69 & -0.80 & -0.80 & 0.25 & 0.25 \\
\hline
\rule[-1ex]{0pt}{3.5ex} C & 37 & 3.00 & 99.9 & 99.9 & 0.16 & 0.18 & -0.66 & -0.66 & 0.41 & 0.41 \\
\hline
\end{tabular}
\end{center}
\end{table}

\begin{figure}[t]
    \centering
    \includegraphics[width=0.65\linewidth]{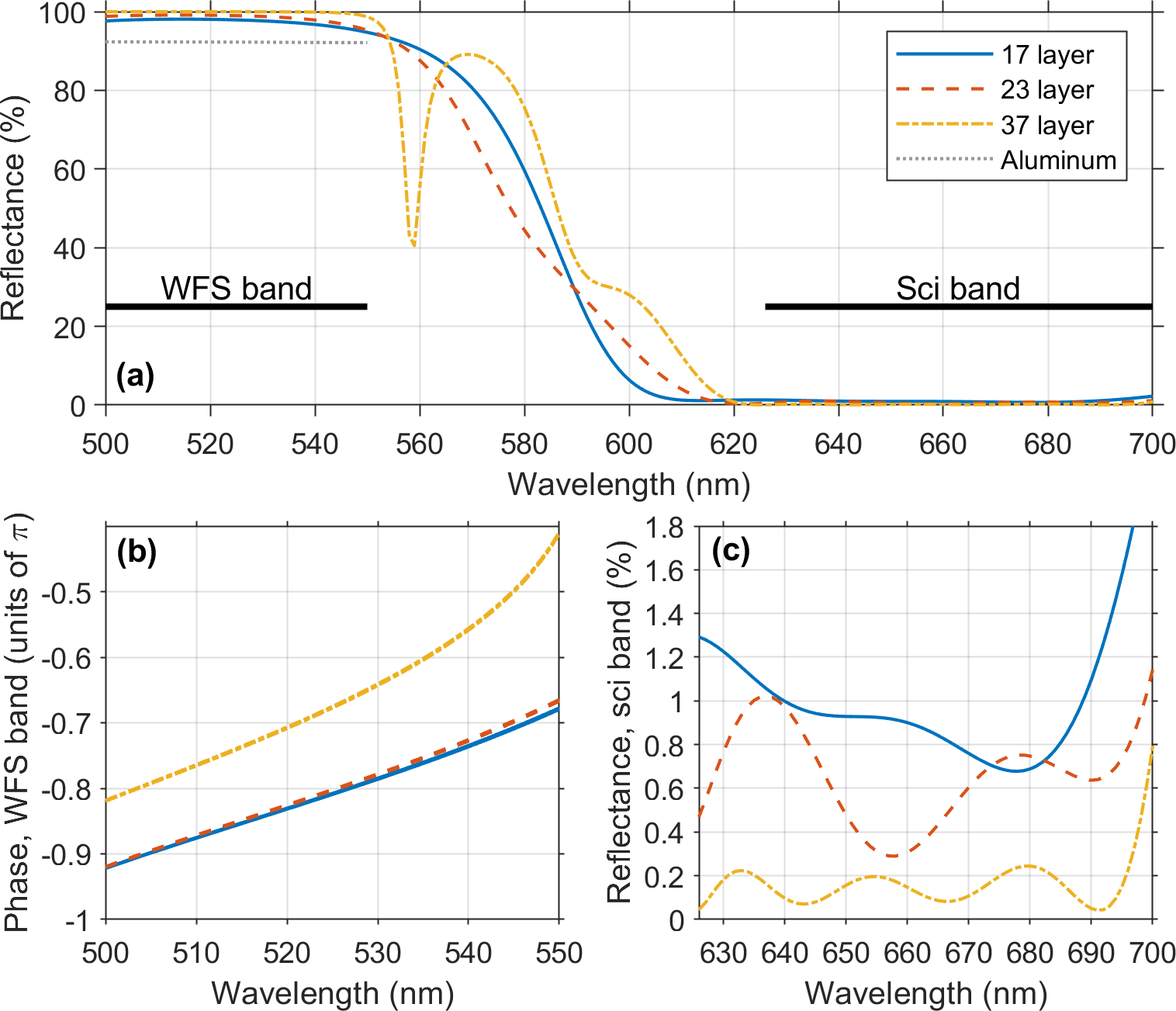}
    \caption{Dichroic coating properties using alternating layers of HfO$_2$ and SiO$_2$. (a)~The reflectance of the dichroic coatings. A 100~nm thick layer of bare Al is included for comparison in the WFS band. (b)~Phase of the reflected light from the dichroic coatings in the WFS band. (c)~The reflectance of the dichroic coatings in the science band (i.e. same as (a), but with different axis scales for emphasis).}
    \label{fig:nasrat_coatings}
\end{figure}

A key aspect of the DPM design is to match the phases of the light reflected from the occulter and dichroic coating as well as possible. To study these requirements, we produced a set of three coating options using alternating layers of hafnium oxide (HfO$_2$) and silicon dioxide (SiO$_2$). The properties of the coatings are given in Table~\ref{tab:coating_parameters} and Fig.~\ref{fig:nasrat_coatings} over the relevant wavelength ranges. The coatings labeled `A', `B', and `C' have 17, 23, and 37 layers, respectively. As expected, using a larger number of layers improves the transmission in the science band and reflectance in the WFS band, but also impacts the phase of the reflected light. All of the coatings give better reflectance than bare Al in the WFS band and $\lesssim$1\% reflectance in the science band. 

\begin{figure}[t]
    \centering
    \includegraphics[width=0.69\linewidth]{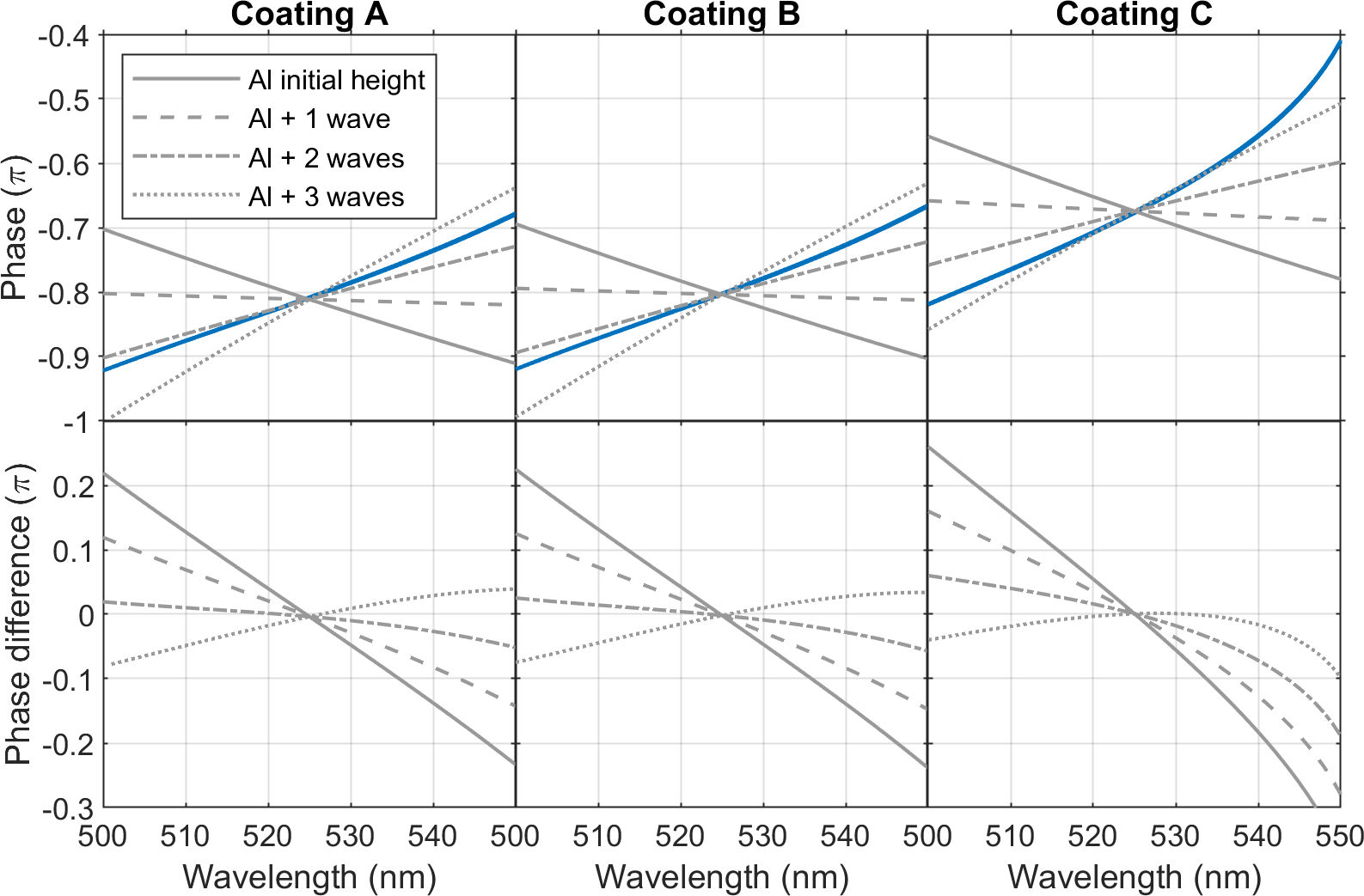}
    \includegraphics[trim={0 -3.0cm 0 0},width=0.3\linewidth]{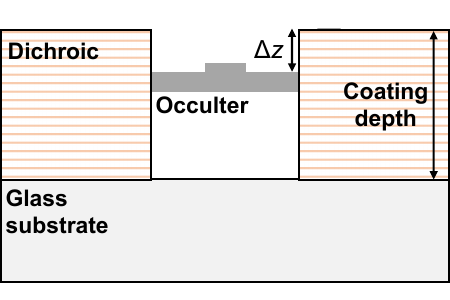}
    \caption{Comparison between reflected phase from dichroic coatings (blue lines) and the bottom layer of the Al occulter (gray lines). The initial positions of the Al surfaces are 300, 301, and 318~nm above the top of the dichroic coatings, respectively. Lowering the Al occulter with respect to the top of the dichroic improves phase matching over the whole wavelength range. The gray lines are half-wave steps in height (full-wave in optical path) at 525~nm. The occulter locations and phase difference ranges are given in Table~\ref{tab:coating_phases}. The diagram at the right shows the definition of $\Delta z$.}
    \label{fig:Al_depth_comparison}
\end{figure}

Figure~\ref{fig:Al_depth_comparison} compares the phase of the light reflected from the dichroics with the Al occulter. We initially set the height to the Al surface above the top of the dichroic coating to match the phase of the light reflected from both regions at the central wavelength of 525~nm; the initial positions, $\Delta z$, are respectively -300~nm, -301~nm, and -318~nm for the three coatings, where the top of the dichroic coating is defined as position zero and the negative sign indicates above the top surface. Since the light that reflects from the dichroic has an additional phase shift due to path length difference to the dichroic coating and the interaction with the dichroic layers, the slope of the phase versus wavelength is different between the dichroic and Al regions. Lowering the Al occulter into the dichroic coating (i.e. increasing $\Delta z$) gives a better match between the phases in the Al and dichroic regions over the WFS wavelength range. 

The phase of the light reflected from the Al and dichroic coating regions and the differences between them, $\Delta\psi$, are shown in Fig.~\ref{fig:Al_depth_comparison} and given numerically in Table~\ref{tab:coating_phases}. For example, in the case of coatings A and B, the $\Delta\psi$ range is roughly a quarter-wave (0.5$\pi$) for the initial occulter position of 300~nm above the coating and an eighth-wave (0.25$\pi$) when the occulter is 38~nm above the coating. Moving the occulter down by another wave (i.e. by a $\Delta z$ step of 525~nm/2 = 262.5~nm) makes the $\Delta\psi$ range 0.07-0.08$\pi$, which is the best possible phase match over the full wavelength range, but requires $\Delta z$~=~225~nm (i.e. that the top of the Al is 225~nm lower that the top of the dichroic coating). The coating C uses more layers and a larger overall depth (3~$\mu$m for coating C vs. 1.4~$\mu$m for coating A) and therefore the occulter would need to be deeper ($\Delta z$~=~470~nm) into the coating to provide the ideal phase match. 

While the simplest manufacturing approach would likely be to deposit the metallic occulter on top of the dichroic coating, requiring broadband phase matching would significantly complicate the DPM manufacturing. Since the transmission of the Al needs to be $<$10$^{-4}$, the minimum thickness is approximately 70~nm, and thus the metallic occulter will need to be at least partially below the dichroic coating for solutions where the position is $>$-70~nm. In the performance modeling section (Section~\ref{sec:performance}), we investigate the impact of the occulter $\Delta z$ position on the WFS sensitivity.

\begin{table}[t]
\caption{Occulter locations and corresponding range of phase differences between the reflected light from the occulter and dichroic, $\Delta\psi$, over the WFS spectral band. Initially the Al surface is above (negative $\Delta z$) the top of the dichroic coating. Each design adds one wave of propagation at the central wavelength of 525~nm in the Al region (i.e. physically lowers the Al occulter towards the substrate). } 
\label{tab:coating_phases}
\begin{center}       
\begin{tabular}{|c|c|c|c|c|} 
\hline
\rule[-1ex]{0pt}{3.5ex}  &  Coating: & A & B & C \\
\hline
\rule[-1ex]{0pt}{3.5ex} \multirow{2}{6em}{Al initial} & $\Delta z$ (nm) & -300 & -301 & -318 \\
\cline{2-5}
\rule[-1ex]{0pt}{3.5ex}   & $\Delta\psi$ range ($\pi$) & 0.45 & 0.46 & 0.63 \\
\hline
\rule[-1ex]{0pt}{3.5ex} \multirow{2}{6em}{Al + 1 wave} & $\Delta z$ (nm) & -37.5 & -38.5 & -55.5 \\
\cline{2-5}
\rule[-1ex]{0pt}{3.5ex}   & $\Delta\psi$ range ($\pi$) & 0.26 & 0.27 & 0.44 \\
\hline
\rule[-1ex]{0pt}{3.5ex} \multirow{2}{6em}{Al + 2 waves} & $\Delta z$ (nm) & 225 & 224 & 207 \\
\cline{2-5}
\rule[-1ex]{0pt}{3.5ex}   & $\Delta\psi$ range ($\pi$) & 0.071 & 0.082 & 0.25 \\
\hline
\rule[-1ex]{0pt}{3.5ex} \multirow{2}{6em}{Al + 3 waves} & $\Delta z$ (nm) & 487.5 & 486.5 & 469.5 \\
\cline{2-5}
\rule[-1ex]{0pt}{3.5ex}   & $\Delta\phi$ range ($\pi$) & 0.12 & 0.11 & 0.099 \\
\hline
\end{tabular}
\end{center}
\end{table}

We envision several extensions of the dual purpose concept to different coronagraph types and wavefront sensing methods. For example, in future work we will explore adding dichroic coatings and/or phase-shifting features to vector vortex coronagraphs. On the other hand, we are developing an enhancement to the mask design that uses a metamaterial to provide polarization dependent phase shifts of $\pm \lambda_\text{wfs}/4$ with respect to the bottom layer of the metallic occulter, where the sign of the phase shift is determined by the orientation of the metamaterial elements with respect to the incident polarization\cite{Wenger2023}. In that case, the resulting WFS effectively operates similarly to a vector-Zernike WFS\cite{Doelman2019}. In the following sections, we present modeling and initial laboratory results that are aimed at enabling the DPLC scalar and metamaterial varieties as well as other coronagraph types.


\section{Performance modeling} 
\label{sec:performance} 

In this section, we present theoretical modeling of the coronagraph (i.e. science) path and the WFS path for the DPLC design described above.  

\subsection{Coronagraph / ``Science" path}

The best performing Lyot coronagraphs to date in NASA's High Contrast Imaging Testbed (HCIT) facility use two DMs in series to apodize the stellar beam\cite{Seo2019}. In this approach, conventional focal plane wavefront sensing and control techniques\cite{Groff2015} are used to find the DM settings that provide high contrast in the desired field-of-view and wavelength range. Here we assume two DMs with 46 actuators across the beam, an inter-actuator pitch of 1~mm, and a 1~m separation between the DMs (similar to HCIT testbeds and the RST coronagraph instrument). 

We consider two types of solutions that use the same DPM: (1) blind search modes and (2) spectroscopy modes. The first is optimized to create high contrast on both sides of the star, whereas the second concentrates on a more localized region in the image plane. Using the two-sided image is advantageous for discovering previously unknown exoplanets. On the other hand, using a small region of the image when spectrally characterizing a known exoplanet provides higher throughput and larger bandwidth than the blind search mode. Indeed, the majority of a coronagraph mission observation time is likely spent taking spectra and thus an optimized spectroscopy mode may drastically improve the efficiency and yield of the mission. 

\begin{figure}[t]
    \centering
    \includegraphics[width=0.6\linewidth]{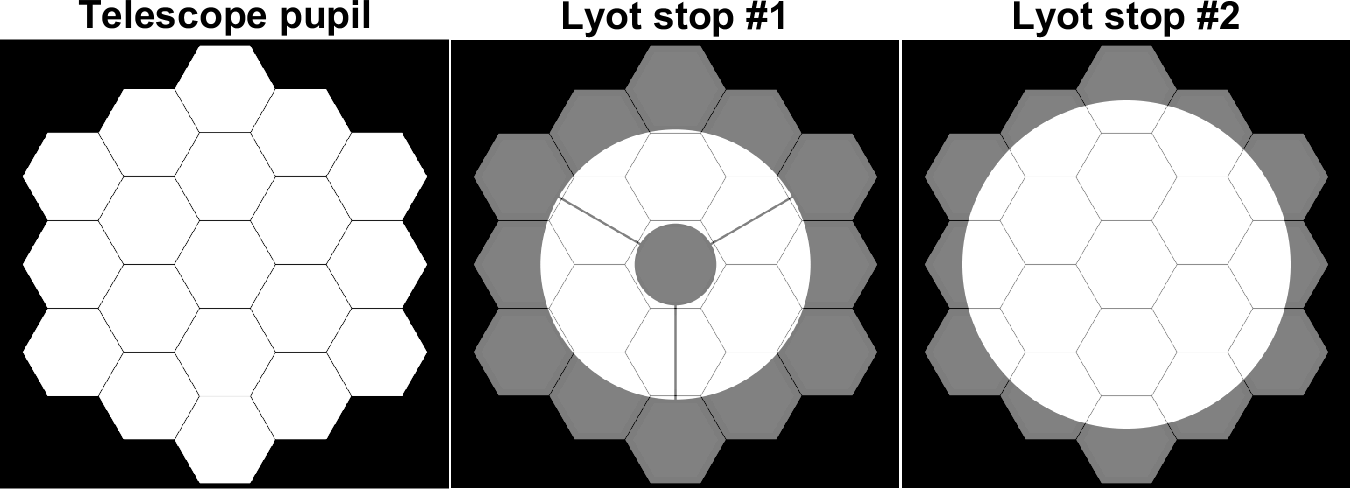}
    \caption{A possible telescope pupil for HWO as well as the Lyot stop designs for the blind-search (Lyot stop \#1) and spectroscopy (Lyot stop \#2) modes. The Lyot stops are overlayed on the telescope pupil.}
    \label{fig:coro_pupils}
\end{figure}

\begin{figure}[t]
    \centering
    \includegraphics[width=0.7\linewidth]{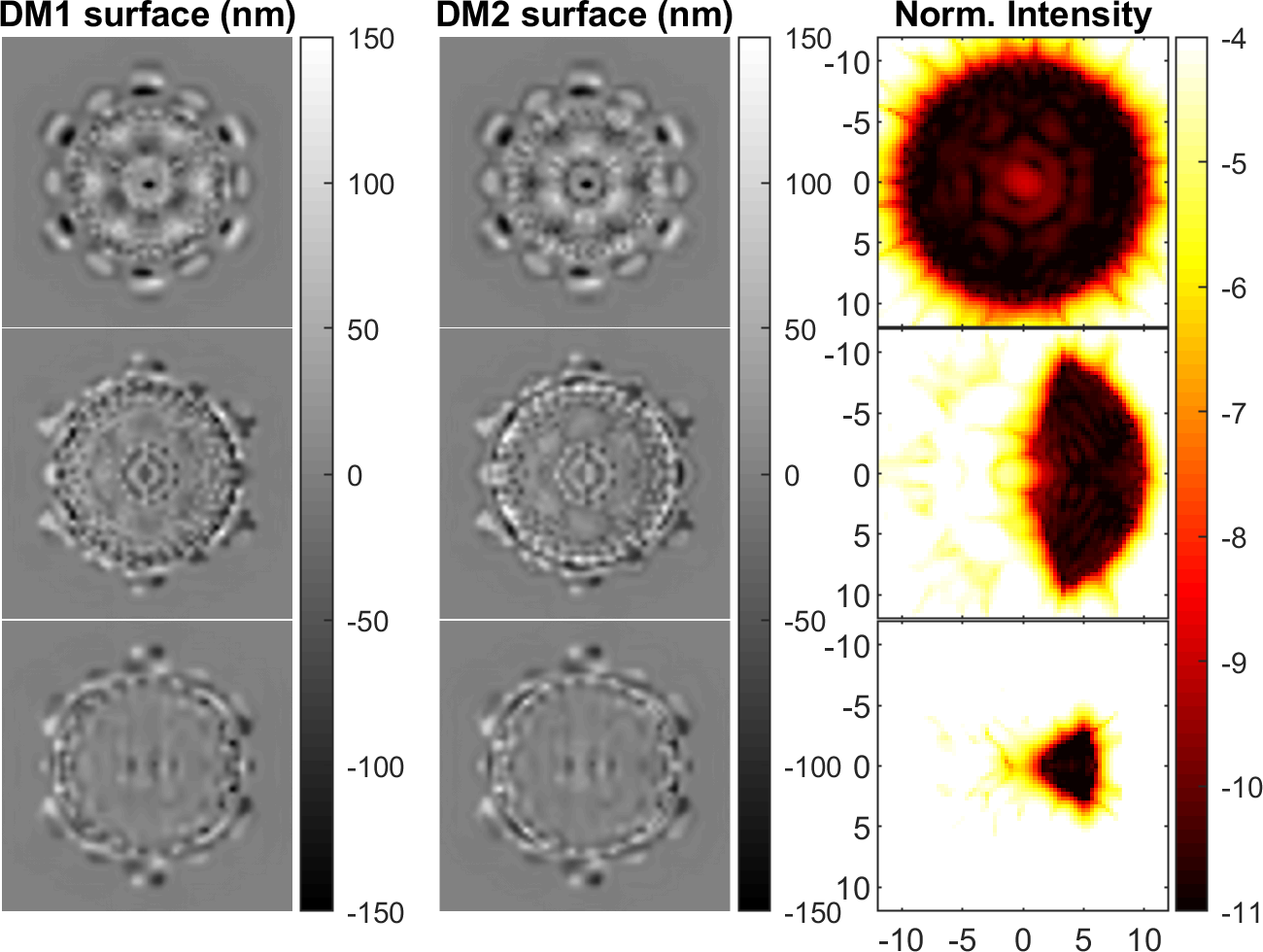}
    \caption{Coronagraph modes optimized for blind-search (top row) and spectroscopy (middle/bottoms rows). Using the same DPM occulter, the DM shapes (columns 1 and 2) are optimized to minimize the normalized intensity (column 3) in specific regions and bandwidths. The coronagraph parameters are quantified in Table~\ref{tab:coro_model}.}
    \label{fig:DMshapes_and_DHs}
\end{figure}

The assumed pupil is shown in Fig.~\ref{fig:coro_pupils} along with the two Lyot stops for the blind-search (Lyot stop \#1) and spectroscopy (Lyot stop \#2) modes. Figure~\ref{fig:DMshapes_and_DHs} shows the optimized DM surface shapes (first and second columns) and the resulting stellar intensity assuming a point source. The top row is the solution for the blind-search mode, which uses Lyot stop \#1, a two-sided dark zone, and a 10\% spectral bandwidth. The bottom two rows are potential spectroscopy modes that use Lyot stop \#2, a one-sided dark zone, and a 20\% spectral bandwidth. Here, we show the 20\% bandwidth for the spectroscopy mode in these models for demonstration purposes despite the coatings presented above being optimized for the 10\% bandwidth case. In the future, we will explore coating designs that accommodate the 20\% bandwidth coronagraph modes as well.

\begin{figure}[t]
    \centering
    \includegraphics[width=0.7\linewidth]{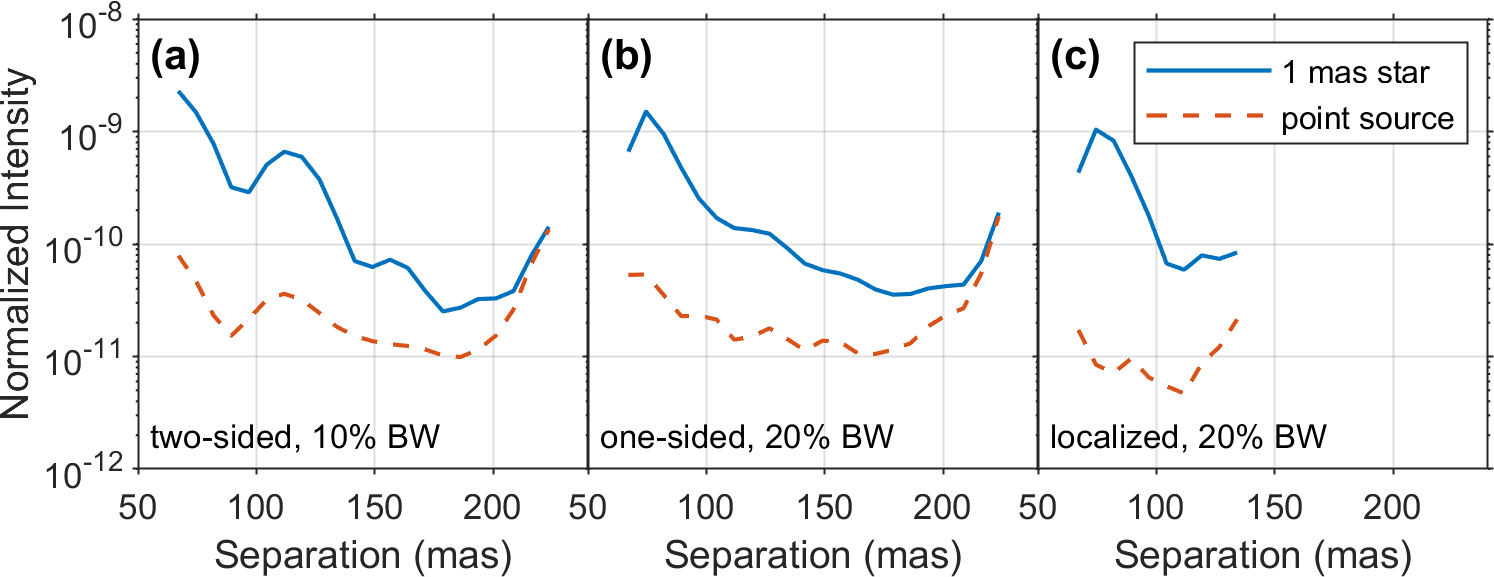}
    \caption{Azimuthal average normalized intensity for the coronagraph modes shown in Fig.~\ref{fig:DMshapes_and_DHs}, including the impact of the 1-mas diameter star. The assumed telescope diameter is 6~m. }
    \label{fig:normI_radialProfiles}
\end{figure}

Figure~\ref{fig:normI_radialProfiles} shows the azimuthal average (i.e. radial profile) of the normalized intensity in the three coronagraph modes. The solid line shows the predicted normalized intensity for a star that is 1~milliarcsecond (mas) in diameter. The large increase between a point source model and a star of realistic size stems from the high sensitivity of these coronagraph designs to small tip/tilt errors, which tends to be a primary weakness of simple Lyot coronagraphs. Our team is actively exploring ways to mitigate this effect by directly constraining the low-order sensitivities in the coronagraph design process, which could significantly improve the contrast at small separations\cite{Riggs2019}. 

Relevant coronagraph model parameters are given in Table~\ref{tab:coro_model}, including the resulting surface height RMS on each DM, throughput, and normalized intensity at different separations. Generally speaking, the spectroscopy modes (one-sided dark holes) allow larger bandwidth, reduce the required DM surface variations, and improve the throughput over the blind-search settings. However, for a 1-mas star, the expected normalized intensity at 3-4~$\lambda/D$ (67-89~mas) separations is $\sim$10$^{-9}$ in all three coronagraph modes. Improving the contrast at $<$100~mas separations remains a primary challenge for the coronagraph design aspect of this work. Nonetheless, laboratory demonstrations of similar coronagraphs in the blind-search\cite{Seo2019} and spectroscopy\cite{Allan2023_SPIE} modes provide confidence that these contrast levels can be reached in practice. 

\begin{table}[t]
\caption{Coronagraph model parameters. $^*$Occulter radius. $^\dagger$Mean normalized intensity in correction region assuming a 1-mas star. } 
\vspace{-5mm}
\label{tab:coro_model}
\begin{center}       
\begin{tabular}{|c|c|c|c|c|c|c|c|c|c|} 
\hline
\rule[-1ex]{0pt}{3.5ex}  Type & $\Delta\lambda/\lambda$ & $R_\text{OCC}^*$ & \multicolumn{2}{|c|}{Corrected region} & \multicolumn{2}{|c|}{RMS surface height} & Throughput & \multicolumn{2}{|c|}{Intensity$^\dagger$ ($\times10^{-9}$)} \\
\hline
\rule[-1ex]{0pt}{3.5ex}   & \%  & $\lambda/D$ & $R$ ($\lambda/D$) & Ang. ($^\circ$) & DM1 (nm) & DM2 (nm) & \%  & 3-4~$\lambda/D$ & 4-5~$\lambda/D$ \\
\hline
\rule[-1ex]{0pt}{3.5ex} Search & 10 & 2.8 & 10 & 180 & 27.0 & 27.6 & 10.7 & 1.2 & 0.4 \\
\hline
\rule[-1ex]{0pt}{3.5ex} Spec. & 20 & 2.8 & 10 & 140 & 22.7 & 24.8 & 19.3 & 1.3 & 0.2 \\
\hline
\rule[-1ex]{0pt}{3.5ex} Spec. & 20 & 2.8 & 6 & 60 & 17.0 & 17.8 & 23.2 & 1.1 & 0.2 \\
\hline
\end{tabular}
\end{center}
\end{table}

\subsection{Wavefront sensor path}

\subsubsection{Impact of the dichroic phase}

A DPM with a perfect match in the reflected amplitude and phase between the occulter and the dichroic is a conventional Zernike WFS. Following the theoretical analysis in the appendix of Ruane et al. (2020)\cite{Ruane2020_JATIS} and references therein\cite{Wallace2011,NDiaye2013_ZELDA}, the intensity at the WFS camera $I_\text{Z}(x,y) = |E_\text{c}(x,y)|^2$ may be expanded as 
\begin{equation}
    E_\text{c}(x,y) = \left[E_\text{p}(x,y) - b(x,y)\right] + b(x,y)e^{i\theta}, 
    \label{eqn:interference}
\end{equation}
where $E_\text{p}(x,y)$ is the re-imaged pupil field, $b(x,y)$ is the so-called reference wave, and $\theta$ is the phase shift provided by the WFS dimple (i.e. the top layer in the DPM occulter relative to the bottom layer). Heuristically, $b(x,y)$ is a low-pass filtered version of $E_\text{p}(x,y)$ containing spatial frequencies up to $\hat{d}=d/(\lambda F^\#)$ cycles across the pupil, where $d$ the phase dimple diameter. Expanding the complex-field in the pupil of $E_\text{p}(x,y) = A(x,y)e^{i\phi(x,y)}$, where $A(x,y)$ is the amplitude and $\phi(x,y)$ is the phase, the pupil intensity at the WFS camera with the phase dimple aligned to the center of the focused beam is
\begin{equation}
    I_\text{Z} = A^2 + 2 b^2(1-\cos\theta) + 2 A b \chi(\phi,\theta),
    \label{eqn:IZ}
\end{equation}
where $\chi(\phi,\theta) = \sin\phi \sin\theta - \cos\phi (1-\cos\theta)$, $b$ is assumed to be real-valued, and we have dropped the $(x,y)$ transverse coordinates for brevity.

\begin{figure}[t]
    \centering
    \includegraphics[width=0.75\linewidth]{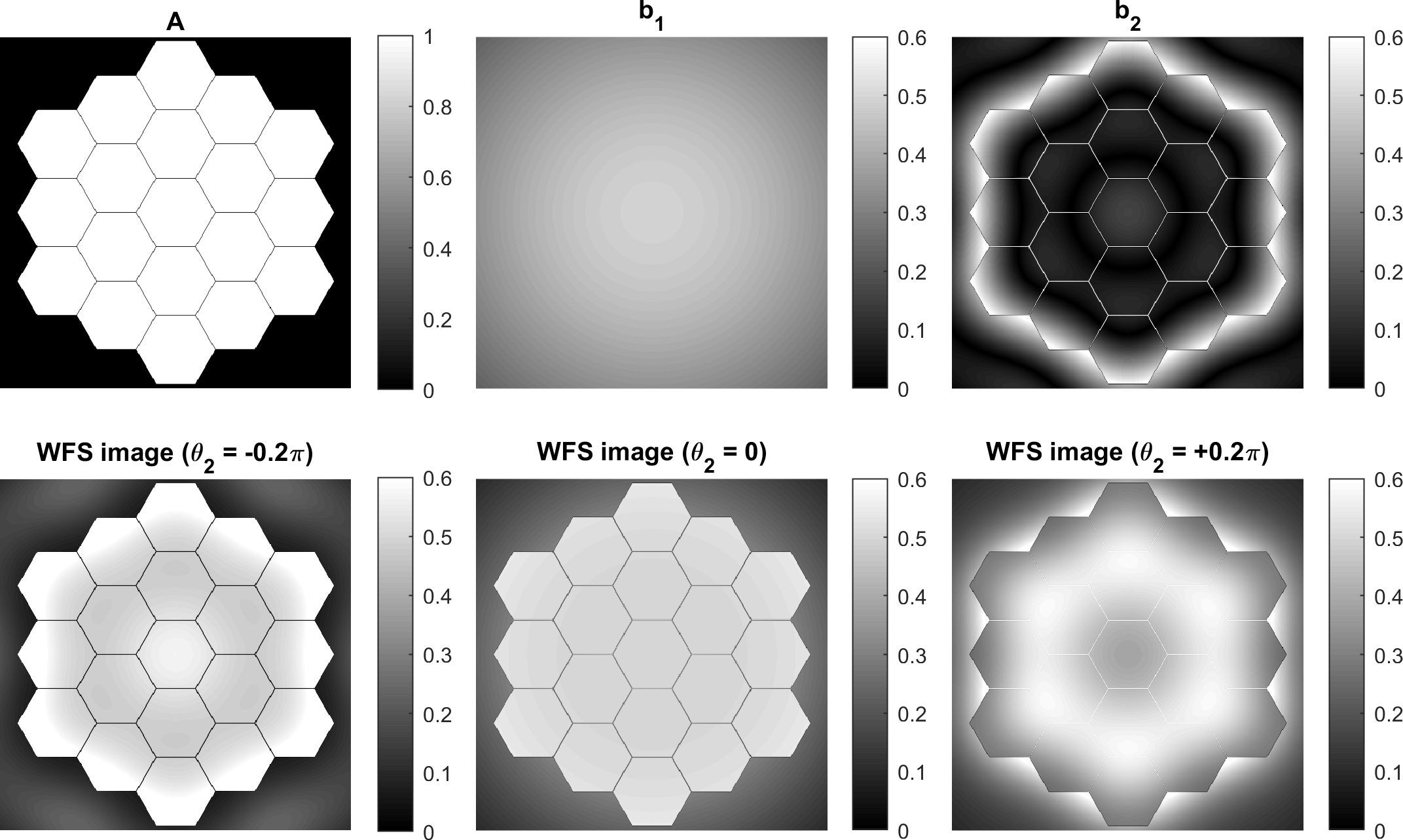}
    \caption{The components of the WFS image. The top row shows the $A$, $b_1$, and $b_2$ amplitudes for the DPM design in Table~\ref{tab:fpm_design}. The bottom row shows approximated monochromatic WFS images at 500~nm, 525~nm, and 550~nm assuming respective phase shifts in the dichroic of -0.2$\pi$, 0, and +0.2$\pi$ (i.e. similar phase shifts as the case with the occulter located 300~nm above the coating surface shown for coating `A' in Fig.~\ref{fig:Al_depth_comparison}).  }
    \label{fig:WFSpupils}
\end{figure}

Adding the effect of the dichroic is a simple extension to Eqn.~\ref{eqn:interference}. However, we need to consider two cases. The first case is where the spatial mode of $\phi$ is low-order and therefore would be spatially filtered out of the part of the field that passes through the dichroic. Thus, Eqn.~\ref{eqn:interference} becomes: 
\begin{equation}
    E_\text{c} = \left[E_\text{p} - b_1 - b_2\right] + b_1 e^{i\theta_1} + b_2 e^{i\theta_2}, 
    \label{eqn:intereference_LO}
\end{equation}
where $b_1$ and $\theta_1$ are the conventional Zernike WFS reference wave and phase shift and $b_2$ and $\theta_2$ are the equivalents for the dichroic region. In the second case, the $\phi$ is a high-order spatial mode and is not filtered from the $b_2$ term and
\begin{equation}
    E_\text{c} = \left[E_\text{p} - b_1 - b_2e^{i\phi}\right] + b_1 e^{i\theta_1} + b_2e^{i\phi} e^{i\theta_2}.
    \label{eqn:intereference_HO}
\end{equation}
To help visualize these terms, the top row of Fig.~\ref{fig:WFSpupils} shows the $A$, $b_1$, and $b_2$ amplitudes for the DPM presented above. The bottom row shows the combined intensity for representative parameters at the shorest, central, and longest wavelengths. 

For both the low-order and high-order cases, we write the intensity at the WFS camera as 
\begin{equation}
    I_\text{WFS} = A^2 + I_1 + I_2 + I_3,
    \label{eqn:IWFS}
\end{equation}
where $I_1$, $I_2$, and $I_3$ are three intensity terms. Regardless of the spatial frequency, 
\begin{equation}
    I_1 = 2 b_1^2(1-\cos\theta_1) + 2 A b_1 \chi(\phi,\theta_1),
\end{equation}
which is the same as the original Zernike WFS expression due to the central phase dimple. On the other hand, $I_2$ and $I_3$ differ for low-order and high-order cases.

For low-order wavefronts, the remaining intensity terms become
\begin{equation}
    I_2 = 2 b_2^2(1-\cos\theta_2) + 2 A b_2 \chi(\phi,\theta_2) \text{ and } 
    I_3 = 8 b_1 b_2 \sin\left(\frac{\theta_1}{2}\right)\sin\left(\frac{\theta_2}{2}\right)\cos\left(\frac{\theta_1-\theta_2}{2}\right).
\end{equation}
$I_2$ is the equivalent of $I_1$ for the dichroic region and $I_3$ is a cross term between the two. To relate the change in intensity on the camera to a change in the phase, we find the derivative of Eqn.~\ref{eqn:IWFS} with respect to $\phi$:
\begin{equation}
    \frac{\Delta I_\text{WFS}}{\Delta \phi} = 2 A (b_1\chi^\prime(\phi,\theta_1) + b_2\chi^\prime(\phi,\theta_2)) ,
    \label{eqn:dIdphi_LO}
\end{equation}
where $\chi^\prime(\phi,\theta) = \cos\phi \sin\theta + \sin\phi(1-\cos\theta)$. Following previous work\cite{Guyon2005}, we then determine the relationship for the error in the phase measurement to the shot noise in the pupil image: $\sigma_{\Delta\phi} = \beta / \sqrt{2 I_\text{e}}$, where $\beta$ is the a WFS sensitivity parameter and $I_\text{e}$ is the signal in each image in units of photo-electrons. In general, $\beta$ may be approximated as
\begin{equation}
    \beta = \sqrt{2 I_\text{e}}\left(\Delta \phi / \Delta I_\text{WFS} \right) \sigma_{\Delta I} , 
    \label{eqn:beta}
\end{equation}
where $\sigma_{\Delta I}$ is the uncertainty in $\Delta I_\text{WFS}$. Substituting Eqn.~\ref{eqn:dIdphi_LO},
\begin{equation}
    \beta = \frac{\sqrt{f_\text{WFS}}}{\sqrt{f_A}\left(\sqrt{f_{b1}}\;\chi^\prime(\phi,\theta_1) + \sqrt{f_{b2}}\;\chi^\prime(\phi,\theta_2) \right)},
\end{equation}
where we've introduced intensity-fraction parameters $f_A$, $f_{b1}$ $f_{b2}$, and $f_\text{WFS}$, such that $A^2 = f_A I_\text{e}$, $b_1^2 = f_{b1} I_\text{e}$, $b_2^2 = f_{b2} I_\text{e}$, and $I_\text{WFS} = f_\text{WFS} I_\text{e}$. In these terms, $\sigma_{\Delta I} = \sqrt{2f_\text{WFS} I_\text{e}}$.

For high-order aberrations, the solution takes the same form as Eqn.~\ref{eqn:IWFS}, except 
\begin{equation}
    I_2= 2 b_2^2(1-\cos\theta_2) + 2 A b_2 \left( \cos\theta_2-1 \right) \text{ and }
    I_3 = 8 b_1 b_2 \sin\left(\frac{\theta_1}{2}\right)\sin\left(\frac{\theta_2}{2}\right)\cos\left(\frac{\theta_1-\theta_2}{2} - \phi\right).
\end{equation}
Again, taking the derivative of Eqn.~\ref{eqn:IWFS}:
\begin{equation}
    \frac{\Delta I_\text{WFS}}{\Delta \phi} = 2 A b_1 \chi^\prime(\phi,\theta_1) + 8 b_1 b_2 \sin\left(\frac{\theta_1}{2}\right)\sin\left(\frac{\theta_2}{2}\right)\sin\left(\frac{\theta_1-\theta_2}{2} - \phi\right). 
    \label{eqn:dIdphi_HO}
\end{equation}
The expression for $\beta$ in this case can be derived by substituting Eqn.~\ref{eqn:dIdphi_HO} into Eqn.~\ref{eqn:beta}. 

\begin{figure}[t]
    \centering
    \includegraphics[width=0.7\linewidth]{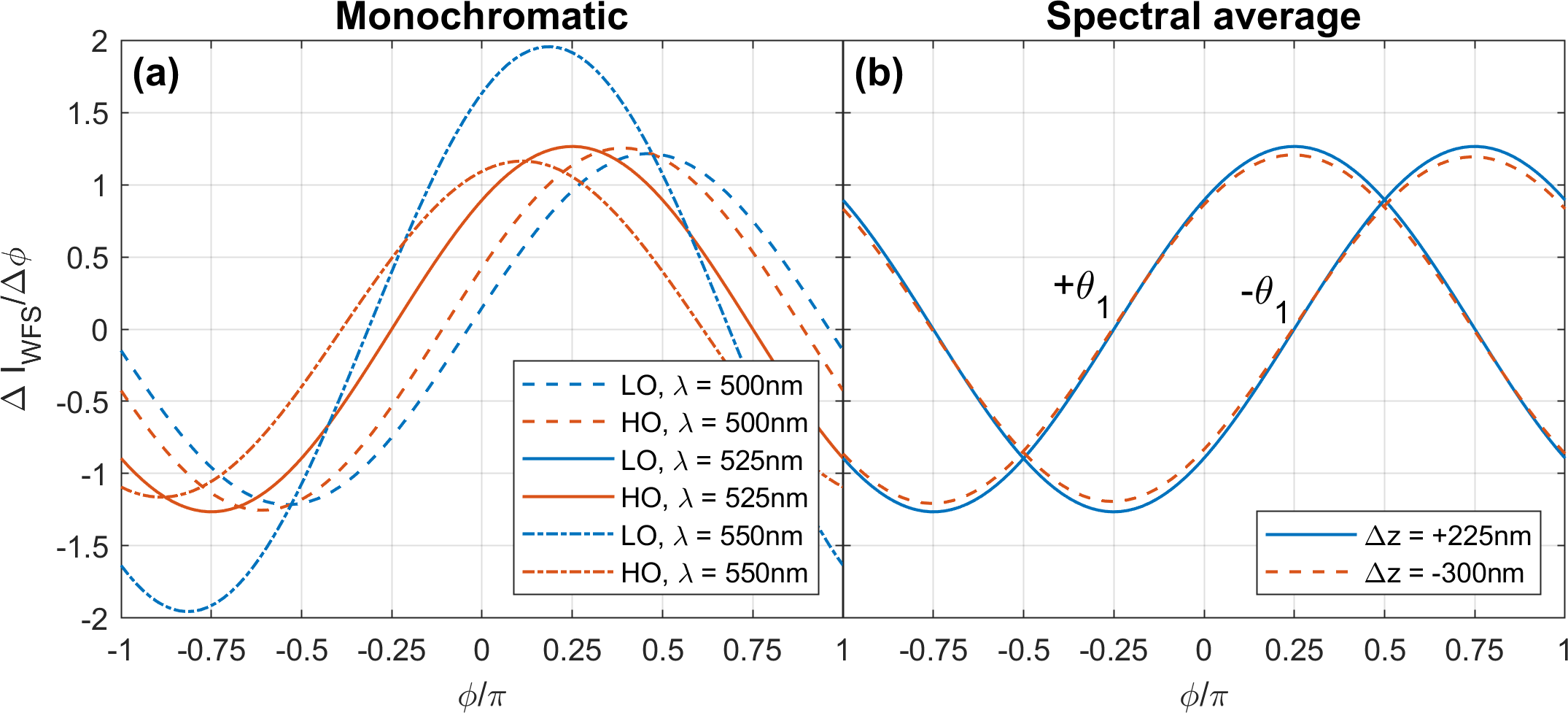}
    \caption{(a)~The WFS response, $\Delta I/\Delta\phi$, for the low-order (LO) and high-order (HO) spatial modes at three wavelengths assuming the phase differences in Fig.~\ref{fig:Al_depth_comparison} for coating A. (b)~The spectral average of the WFS response over the 500-550~nm band and two occulter positions: optimally in-phase ($\Delta z$~=~+225~nm) and offset by three waves ($\Delta z$~=~-300~nm). We also show the case where $\theta_1$ has the opposite sign. }
    \label{fig:dIWFSdPhi}
\end{figure}

Figure~\ref{fig:dIWFSdPhi} shows the WFS response assuming $f_A = 1$, $f_{b1} = 0.2$, $f_{b2} = 0.4$, and $f_\text{Z} = 0.5$, which are representative of a point near the edge of the pupil where $b_2$ is most significant (see Fig.~\ref{fig:WFSpupils}). In Fig.~\ref{fig:dIWFSdPhi}a, we compare the low-order and high-order cases at three representative wavelengths in the WFS band assuming $\theta_2$ is the reflected phase of coating `A' (see Fig.~\ref{fig:Al_depth_comparison}). The response changes with wavelength and does so differently in the low-order and high-order cases. However, when using broadband light, we effectively get the average WFS response which is similar for the low-order and high-order cases and insensitive to the wavelength dependent changes because they are relatively small with respect to the typical values of $\Delta I/\Delta \phi$. In Fig~\ref{fig:dIWFSdPhi}b, we show the spectral average of the WFS response using the high-order model for two cases of $\Delta z$: where $\Delta z$~=~-300~nm and $\Delta z$~=~+225~nm. The former is the case where the occulter is 300~nm above the dichroic surface, which is likely the easiest to manufacture. The latter case has the occulter 225~nm below the top of the dichroic, which is ideal for phase matching at all wavelengths. We conclude from Fig~\ref{fig:dIWFSdPhi}b that the impact of the occulter position on WFS response is relatively minor and confirm that this is true for both positive and negative values of $\theta_1$.

\begin{figure}[t]
    \centering
    \includegraphics[width=0.7\linewidth]{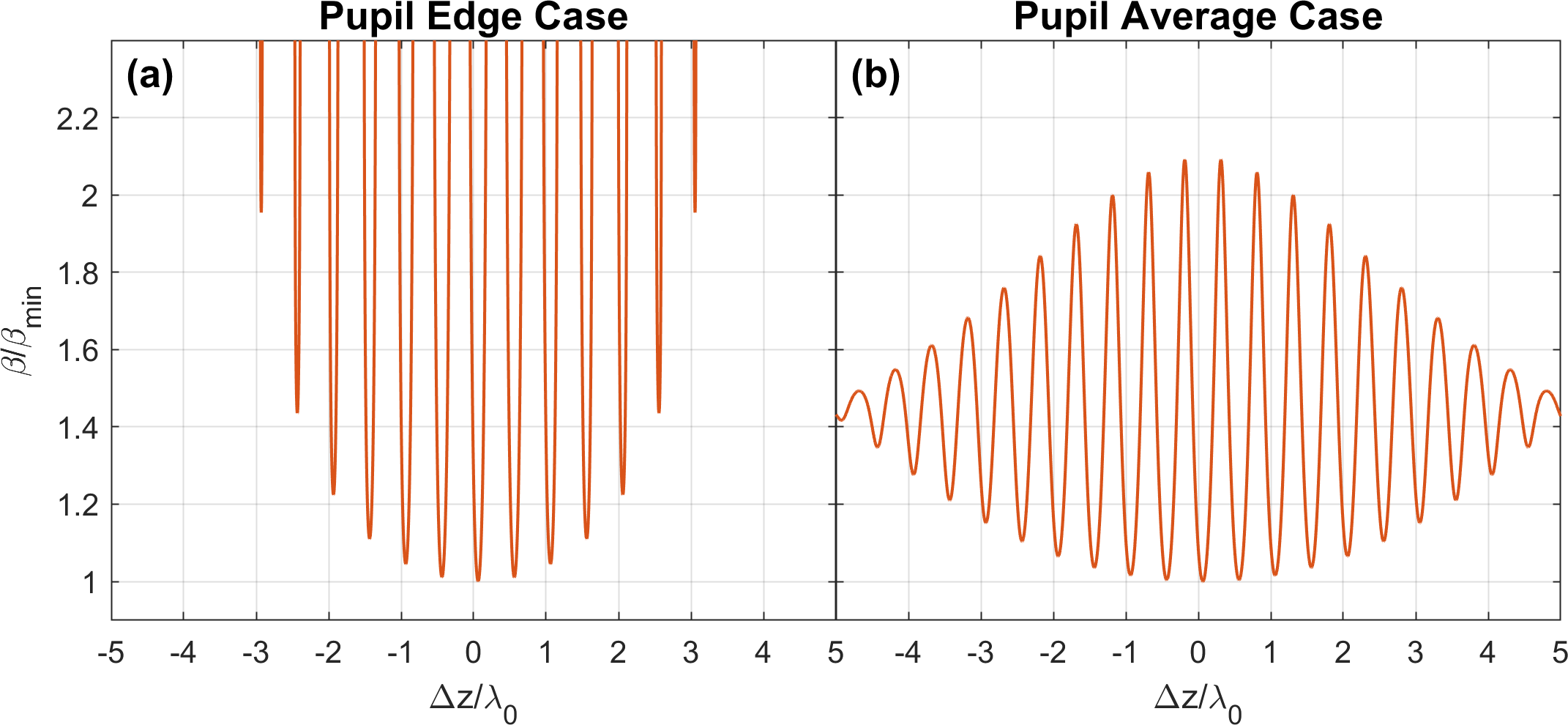}
    \caption{The broadband WFS sensitivity, $\beta$, normalized to its minimum value, as a function of the occulter position with respect to the dichroic surface assuming $\theta_2 = 4\pi\Delta z /\lambda$ and $\phi=0$. }
    \label{fig:beta}
\end{figure}

In terms of the WFS sensitivity, there is a degradation due the phase mismatch between the dichroic and the occulter that depends on the location in the pupil. To demonstrate this, we introduce a simpler model for the dichroic where $\theta_2 = 4\pi\Delta z /\lambda$ (i.e. treating the dichroic as a perfect mirror). Figure~\ref{fig:beta} shows $\beta$ in broadband light (500-550~nm) as a function of $\Delta z$ for two cases representative of an edge region of the pupil ($f_{b1}$~=~0.2, $f_{b2}$~=~0.4) and the average over the pupil ($f_{b1}$~=~0.2, $f_{b2}$~=~0.04). For instance, in the edge case (Fig.~\ref{fig:beta}a), the sensitivity degrades by roughly 10\% when $\theta_2$ changes by three waves (i.e. moves by three fringes in Fig.~\ref{fig:beta}a) from the position with the best (minimum) sensitivity. The impact is significantly smaller in other parts of the image where $b_1$ is smaller (see Fig.~\ref{fig:beta}b). 

We also use Fig.~\ref{fig:beta} to derive phasing requirements based on a maximum allowable value of $\beta$. The precise requirements will depend on how many waves away from the optimal position and the tolerance on $\beta$ versus the pupil position. For our future DPMs, we adopt a $\Delta z$ accuracy requirement of $<\lambda_\text{wfs}/10$. 

\subsubsection{Impact of the initial wavefront}

\begin{figure}[t]
    \centering
    \includegraphics[width=\linewidth]{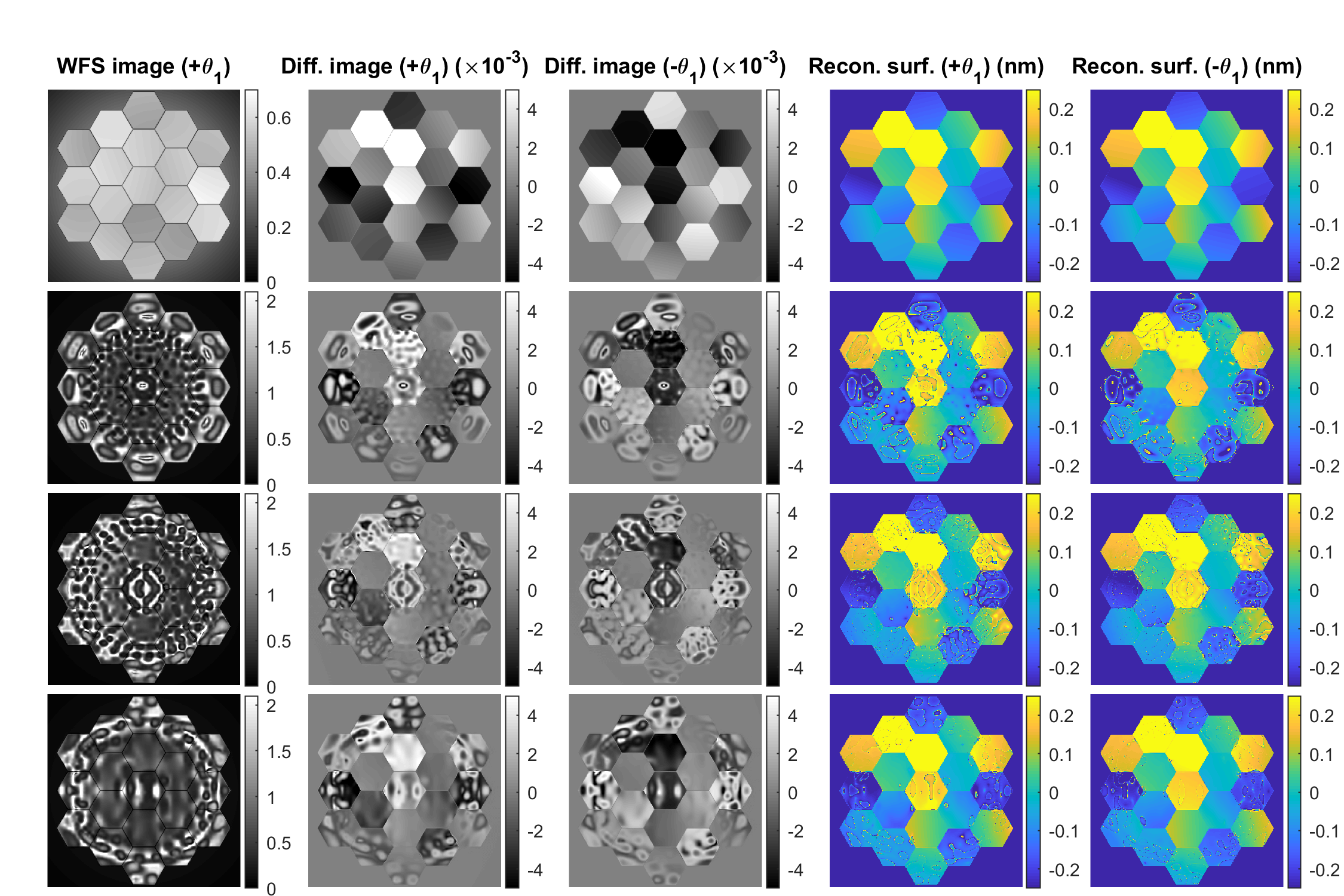}
    \caption{Simulated WFS images for an example scenario where initial telescope segment phasing errors (5~nm~RMS pistons, 1/50~$\lambda/D$~RMS in tip and tilt) change by 260~pm~RMS in piston and 10$^{-3}$~$\lambda/D$~RMS in tip and tilt. The rows respectively correspond to cases with flat DMs (top row) and the DM settings shown in Fig.~\ref{fig:DMshapes_and_DHs} (rows 2-4). The first column shows the initial WFS images. Columns 2-3 are the difference images between the two wavefront states with positive and negative values for $\theta_1$ (i.e. $\theta_1$~=~$\pm\pi/2$ at 525~nm). Columns 4 and 5 are the corresponding reconstructed surface errors to columns 2 and 3. }
    \label{fig:WFSsimulations}
\end{figure}

During coronagraph observations, because the DMs are used for apodization in the coronagraph channel, the initial wavefront seen by the WFS has large spatial variations in both amplitude (i.e. $A\ne1$) and phase (i.e. $\phi \ne 0$). We illustrate the impact of this in Fig.~\ref{fig:WFSsimulations} where we've used a simple propagation model (using FALCO\cite{Riggs2018}), assuming a perfectly phase-matched dichroic, but accounting for the DM settings of the coronagraphs set-ups shown in Fig.~\ref{fig:DMshapes_and_DHs}. For demonstration purposes, we assume an initial error in the telescope segment phasing (5~nm~RMS pistons, 1/50~$\lambda/D$~RMS in tip and tilt) and then a small change from that state (260~pm~RMS pistons, 10$^{-3}$~$\lambda/D$~RMS in tip and tilt). Our goal in this example is to measure the small change in the segment positions in the presence of representative segment position errors and DM shapes. The first column in Fig.~\ref{fig:WFSsimulations} shows the WFS intensity images. The top row assumes the DMs are flat. The next three rows are respectively for the three coronagraph set-ups in the rows of Fig.~\ref{fig:DMshapes_and_DHs}. 

The difference between the WFS images taken with each telescope state are shown in columns 2 and 3 of Fig.~\ref{fig:WFSsimulations} assuming the Zernike dimple is +65.6~nm or -65.6~nm with respect to the larger part of the occulter (i.e. $\theta_1$~=~$\pm\pi/2$ at 525~nm). In the case of flat DMs, the signature of the segment phasing errors is clearly visible in the images. However, in the other three cases where the DMs have large deviations from flat, the images have more complex features due to the spatially varying WFS response. 

The last two columns of Fig.~\ref{fig:WFSsimulations} show the reconstructed surface change based on Eqn.~\ref{eqn:dIdphi_LO} where $b_2=0$. With strong variations in $A$ and $\phi$, estimating the correct phase requires a prior estimate of both $A$ and $\phi$, which may be measured in practice using conventional phase retrieval techniques commonly used to calibrate coronagraphs\cite{Ruane2022}. Even when assuming perfect knowledge of the $A$ and $\phi$ at the central wavelength, there are artifacts in the reconstructions due to the wavelength dependence of $A$ and $\phi$ as well as non-linearity near where $\Delta I/\Delta\phi \approx 0$. In the spectroscopy coronagraph modes, the artifacts are somewhat fewer due to the more favorable DM shapes. However, in all cases, the artifacts tend to appear in different regions of the pupil image in the $+\theta_1$ and $-\theta_1$ reconstructions. This effect is our primary motivation for considering a reflective metasurface (see Wenger et al.\cite{Wenger2023}) instead of the scalar dimple. The metasurface provides the $\pm\theta_1$ phase shifts in orthogonal polarizations which can be split into two images on the WFS camera using a similar method as Doelman et al. (2019)\cite{Doelman2019}. 

\section{Manufacturing, metrology, and preliminary lab testing} 
\label{sec:results}

\begin{figure}[t]
    \centering
    \includegraphics[width=\linewidth]{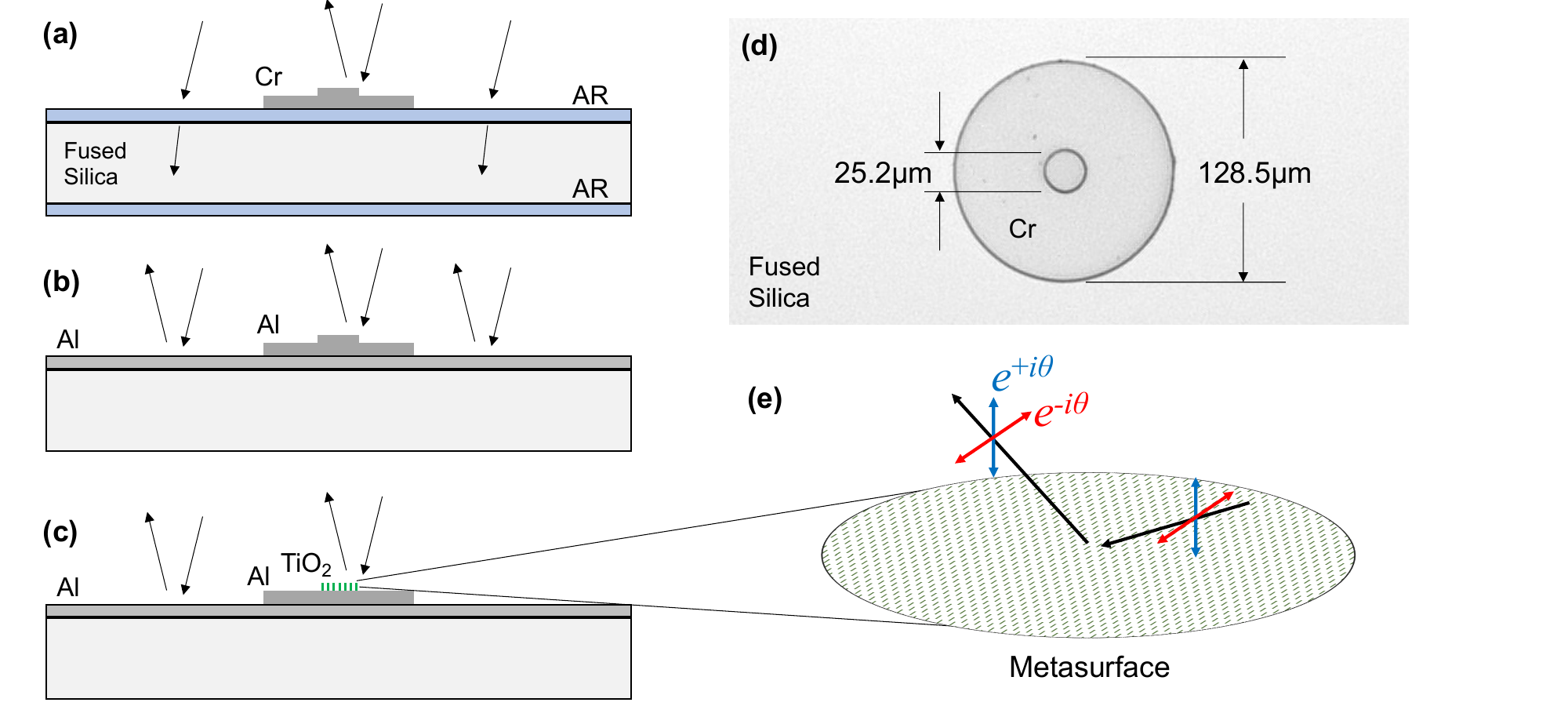}
    \caption{Devices used in development and verification steps. (a)-(c) Masks to verify basic manufacturing and metrology steps as well as coronagraph and WFS functionalities. (d)~Microscope image of the mask in (a). (e)~Diagram of the metasurface concept. The two polarization states (red and blue vectors) each have equal and opposite phase shifts of approximately $\pm\pi$/2 upon reflection (see Wenger et al.\cite{Wenger2023} for more details).}
    \label{fig:fpm_dev}
\end{figure}

We have manufactured a sequence of devices with partial functionality to test specific capabilities in the manufacturing and metrology process. Figure \ref{fig:fpm_dev}a-c show diagrams of our three test masks. The parameters and lessons learned are explained in each case below. 

\subsection{Development mask (a): tiered occulter on AR coated substrate }

The first mask (Fig. \ref{fig:fpm_dev}a) is a two-tiered Cr occulter on a 5~mm thick Fused Silica substrate with a broadband AR-coating (Edmund Optics VIS-NIR; 400-1000~nm) on both the front and back side. The bottom tier of the occulter is 200~nm thick (transmission $<$10$^{-5}$) and has a diameter of 128.5~$\mu$m. The top tier is 77~nm thick and 25.2~$\mu$m in diameter. Several of these occulters were manufactured by Opto-Line International on the 25~mm diameter substrate. Microscope images of these occulters showed very few defects and positioning of the metal layers was within tolerance (see e.g. Fig.~\ref{fig:fpm_dev}d). 

The purpose of this mask was to manufacture a pre-cursory design to the DPM without the out-of-band wavefront sensing functionality. In transmission, it works similarly to the DPM, but the reflected light is spatially filtered at all visible wavelengths and thus only contains the low-order wavefront information at the WFS. In addition, the top-tier is sized to use in-band light for the low-order wavefront sensing. 

\begin{figure}[t]
    \centering
    \includegraphics[height=6.0cm]{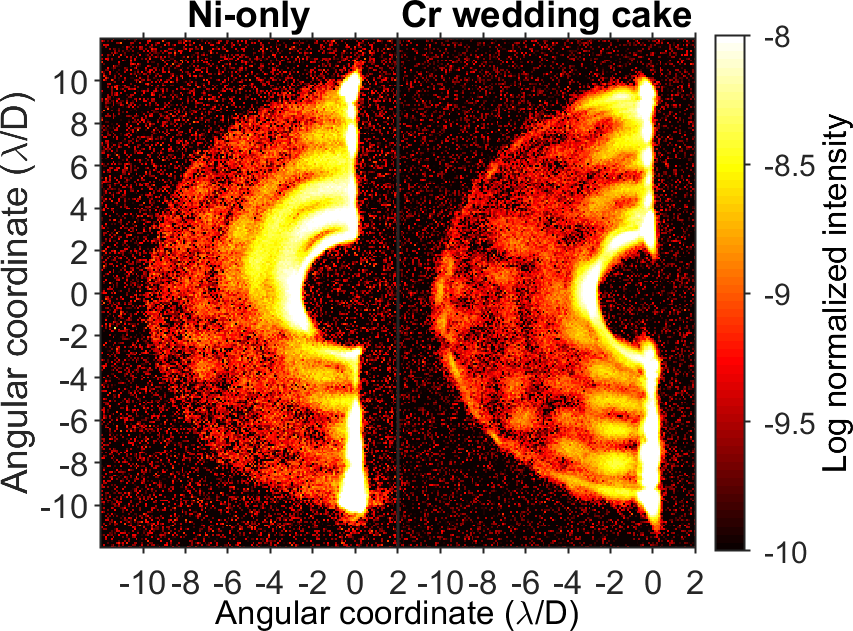}\hspace{0.5cm}
    \includegraphics[height=6.0cm]{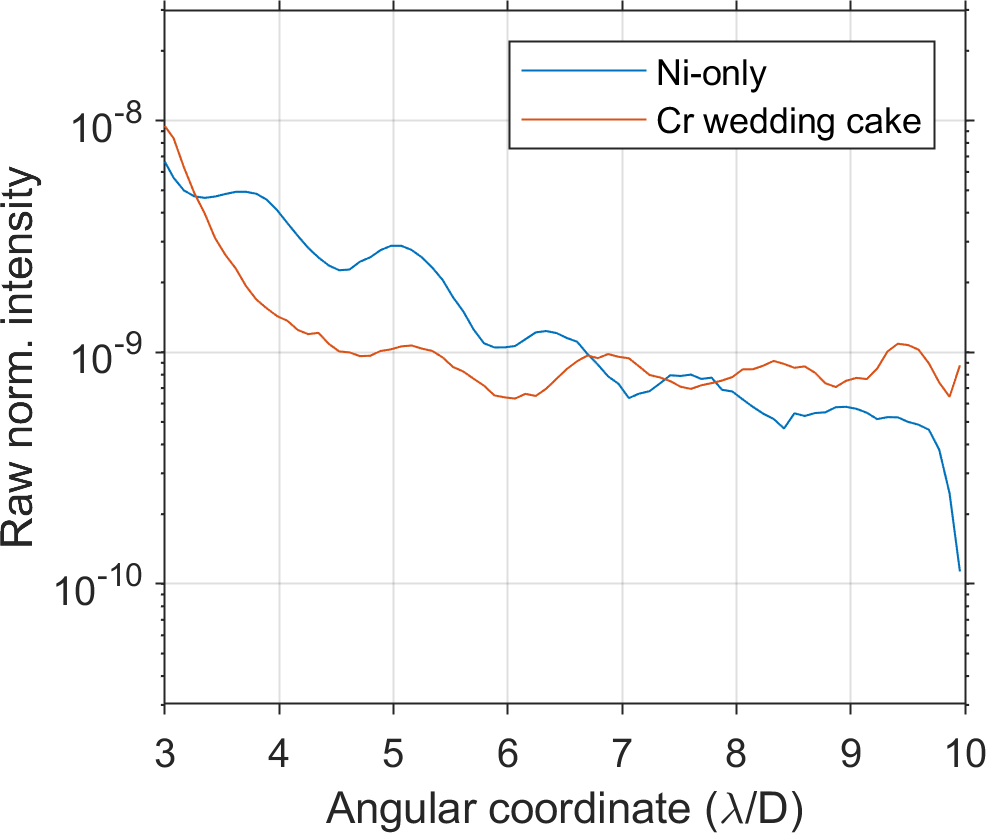}
    \caption{(left panel) Normalized intensity achieved on the DST-2 testbed using a Ni occulter without the phase dimple and the Cr occulter shown in Fig~\ref{fig:fpm_dev}a and \ref{fig:fpm_dev}d. (right panel) The azimuthal average of the normalized intensity images within the field stop region. }
    \label{fig:dst2comparison}
\end{figure}

During the commissioning of the Decadal Survey Testbed \#2 (DST-2)\cite{Meeker2021,Noyes2023_SPIE}, we compared the performance of this mask with the simpler Ni-only occulter used for the commissioning of the original Decadal Survey Testbed (DST-1)\cite{Seo2019}. DST-2 is a new and recently commissioned testbed in NASA's High Contrast Imaging Testbed (HCIT) facility. Using monochromatic light at 660~nm at a circular Lyot stop with 86\% of the full pupil diameter, we created a dark hole on one side of the pseudo-star over 3-10~$\lambda/D$ separations. To create the dark hole, we used one 2040-actuator DM manufactured by Boston Micromachines installed at a re-imaged pupil plane. Figure~\ref{fig:dst2comparison} shows the resulting normalized intensity images. The mean normalized intensities within the correction zone (3-10~$\lambda/D$) with the Ni-only and wedding cake occulters were 1.34$\times$10$^{-9}$ and 1.08$\times$10$^{-9}$, respectively. Since the Ni-only occulter has demonstrated much better contrasts on the DST-1 testbed, we conclude that these experiments are likely not limited by either occulter and that the wedding cake occulter performs at least as well as, and perhaps even better, than the Ni-only occulter in the DST-2. These testbed results, while not at the goal contrast levels of $\sim$10$^{-10}$ for HWO, give us confidence that the manufacturing approach allows $<$10$^{-9}$ contrast. Future testing will be needed to study the contrast performance closer to the $\sim$10$^{-10}$ level. 

\begin{figure}[t]
    \centering
    \includegraphics[width=0.65\linewidth]{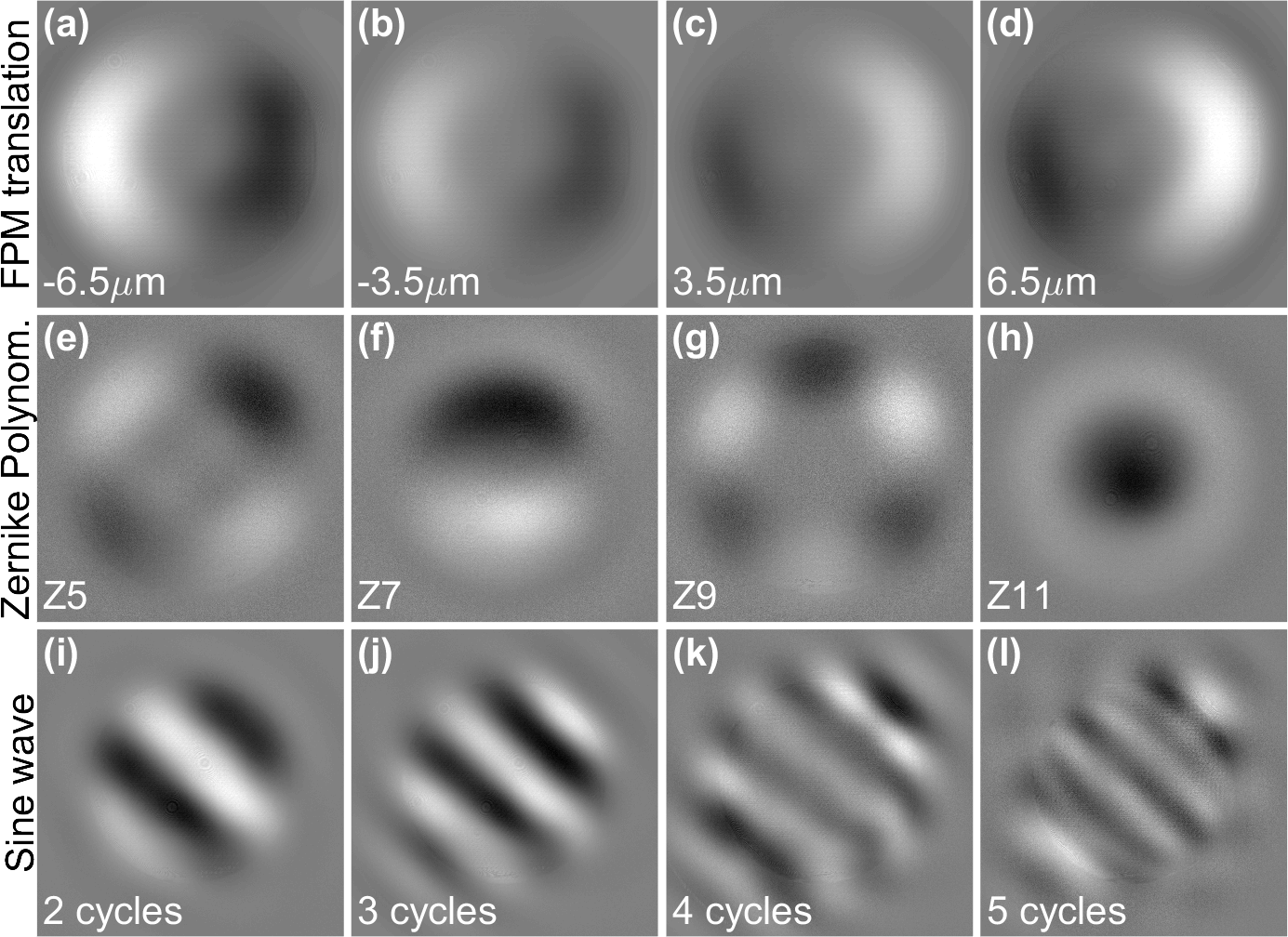}
    \caption{Experimental WFS difference images when applying low-order wavefront errors on the DST-2 testbed in HCIT. The errors were (a)-(d)~light-of-sight via focal plane mask translation, (e)-(h)~Zernike polynomials, and (i)-(l)~Fourier modes of increasing spatial frequency. In (k) and (l), the signal becomes poor due to the spatial filtering effect of the occulter.}
    \label{fig:WFSdata}
\end{figure}

In addition to testing the coronagraph path, we also used this mask to verify the low-order wavefront sensing capability. Figure~\ref{fig:WFSdata} shows the WFS difference images on DST-2 after introducing wavefront errors. We first measured the response to global tip and tilt (i.e. line of sight) errors by translating the focal plane mask (Fig.~\ref{fig:WFSdata}a-d) and showed that we can easily measure $\sim$1~$\mu$m motions (a few \% of $\lambda/D$) in single $\sim$100~ms exposures. We also measured the response to Zernike polynomials (Fig.~\ref{fig:WFSdata}e-h) and Fourier modes (i.e. sine waves; Fig.~\ref{fig:WFSdata}i-l) by applying these patterns to the DM voltages. In the latter, we applied spatial frequencies of 2, 3, 4, and 5 cycles per pupil diameter. As expected the signal is poor beyond 3 cycles because the focal plane mask is not reflective outside of the metallic occulter region. Overall, the low-order wavefront sensor operated as expected. A detailed comparison with theoretical expectation and closed-loop control will the be topic of future work.  

\subsection{Development mask (b): tiered occulter on Al coated substrate }

While ultimately our goal is to fabricate a modified version of the mask in Fig.~\ref{fig:fpm_dev}a, but optimized for out-of-band wavefront sensing, we plan to test the high-order wavefront sensing capability with a known phase shift between the occulter and the surrounding (dichroic) region. Thus, as an intermediary step, we will use an Al wedding cake occulter on top of an Al coated substrate with a known $\Delta z$ offset between the two (see Fig.~\ref{fig:fpm_dev}b). This will allow us to verify the performance models without the complications introduced by the potentially uncertain as-manufactured dichroic phase. This effort is in progress. 

\begin{figure}[t]
    \centering
    \includegraphics[width=\linewidth]{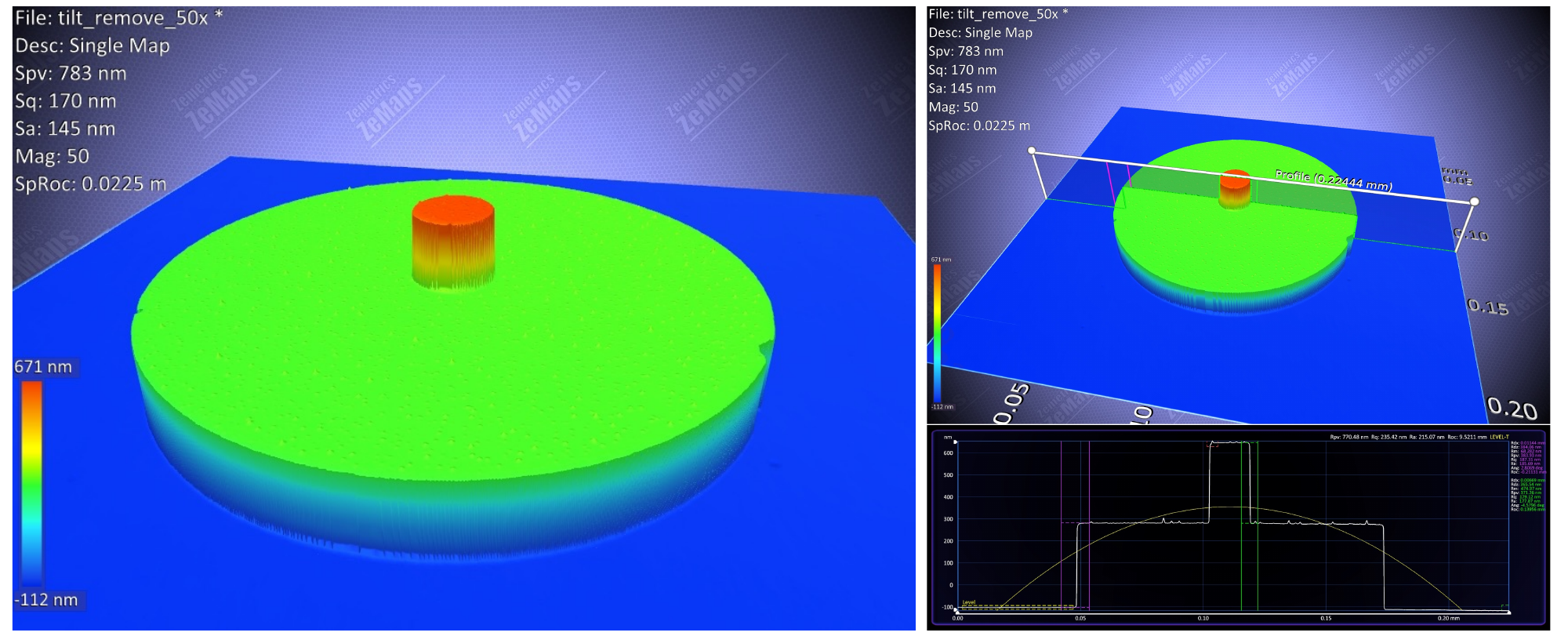}
    \caption{Surface metrology of Al test mask (Fig.~\ref{fig:fpm_dev}b) using the white-light interferometry mode of a Zygo ZeMapper Optical Profiler (screenshot of manufacturer's user interface). The primary purpose of this test was to verify that there was no ambiguity in the absolute step heights, despite having relative phase shifts greater than the wavelength. This is a critical capability for our verifying requirements and characterizing the manufactured DPMs. }
    \label{fig:zemapper}
\end{figure}

In addition to using this style of mask to test the wavefront sensing capability, we also verified the metrology systems we intend to use to measure the phase difference between the dichroic and occulter. To do so, we manufactured a test mask in the Micro Devices Laboratory (MDL) at NASA's Jet Propulsion Laboratory with the spatial scales in Table~\ref{tab:fpm_design} but with step heights of 400~nm. Using a white-light interferometry mode of a Zygo ZeMapper Optical Profiler, we measured the actual step heights to high accuracy without any phase wrapping ambiguities in the data (see Fig.~\ref{fig:zemapper}). This gives us confidence that we will be able to sufficiently verify our final manufacutred DPMs.

\subsection{Development mask (c): Metamaterial phase shifter }

Wenger et al (2023)\cite{Wenger2023} present a metasurface that upon reflected is designed to impart a phase shift of $\pm\pi/2$ where the sign depends on the orientation of the incident polarization with respect to the metasurface structure (see Figs.~\ref{fig:fpm_dev}c and \ref{fig:fpm_dev}e). Splitting the polarizations in the WFS path will effectively allow us to use two WFS images: one with $\theta_1\approx\pi/2$ and one with $\theta_1\approx-\pi/2$. This is achieved by fabricating anisotropic TiO$_2$ pillars on nanometer scales instead of top tier of the occulter. 

There are advantages of having these two simultaneous phase shifts. The principle and primary advantages were explained by Doelman et al.~(2019)\cite{Doelman2019}, who showed that a similar concept can be implemented using geometric phase in liquid crystals and splitting circular polarizations. By combining the data in the two images, they showed that both amplitude and phase changes can be measured simultaneously. Applied to our case, this feature may allow for improved control of the two DMs in the coronagraph instrument.

We also intend to take advantage of the two phase shifts to mitigate artifacts in the WFS reconstructions in cases where there are strong deviations from a uniform plane wave upstream of the WFS. As shown in Fig.~\ref{fig:WFSsimulations}, such artifacts appear in different locations in the image using the two different phase shifts. Closed-loop control with the two WFS images simultaneously promises to mitigate the impact of such artifacts in practice. 

\section{Conclusions} 
\label{sec:conclusions} 

We have presented progress on our quest to develop ``Dual Purpose" coronagraph masks that simultaneously allow high-contrast imaging and high-resolution wavefront sensing using out-of-band starlight. Our design and plan forward takes into account a myriad of system-level considerations unique to space-based coronagraphs and HWO. We provided a detailed example design with a two-tiered Lyot coronagraph occulter on a dichroic coated substrate, which we expect to achieve performance in both terms of contrast and wavefront sensing capability commensurate with observations of the Earth-size exoplanets in the habitable zone of Solar-type stars. Using a combination of numerical simulations and analytical models, we showed the predicted performance of the coronagraph and wavefront sensor taking into account potential phase shifts between the dichroic and occulter as well as the large surface height deviations in the DM shapes of Lyot coronagraphs. Finally, we provided an update on our laboratory testing of masks with partial functionality as part of sequence of verification activities along our way to a complete demonstration of the DPLC in the coronagraph testbeds in NASA's HCIT. 

\acknowledgments 
 
This research was carried out at the Jet Propulsion Laboratory, California Institute of Technology, under a contract with the National Aeronautics and Space Administration (80NM0018D0004).

\bibliography{main} 
\bibliographystyle{spiebib} 

\end{document}